\documentstyle[11pt]{article}

\newcommand{\A}{\mathcal{A}}

\begin{document}

\title{
\bf  { On Torsion Axial Vector and gravitational Energy in Lewis-Papapetrou Space-Time in the Theory of Teleparallel Gravity}}
\author{{A. S. Alofi$^{1,2}$\thanks{%
E-mail: aalofi@kau.edu.sa}\,\, and Ragab M. Gad$^{2,3}$ \thanks{%
E-mail: ragab2gad@hotmail.com} }\\
\newline
{\it $^1$
 Mathematics Department, Faculty of Science, King
Abdulaziz University,}\\
 {\it  21589 Jeddah, KSA}
 \\
 {\it $^2$
 Mathematics Department, Faculty of Science, University of Jeddah,}\\
 {\it  North Jeddah, KSA}
 \\
{\it $3$ Mathematics Department, Faculty of Science, Minia University,}\\
 {\it   61915 El-Minia,  Egypt}
}

\date{\small{}}

\maketitle

\begin{abstract}

The teleparallel version of Lewis-Papapetrou space-time is investigated. For this space-time
the true discussion of the geometric and physical properties is given. We show that the value of
space-like torsion-axial vector does not depend on the choice of tetrad field. Consequently,
the spin procession of the Dirac particle and the corresponding Hamiltonian do.
The energy and momentum densities for such space-time are calculated.
 We  show that when choosing two sets of tetrad fields the dependence and independence of the
 aforementioned quantities on the  tetrad field depend on the Lorentz factor, which links the
 two sets of tetrad fields.

\end{abstract}

\setcounter{equation}{0}
\section{Introduction}

Although General Relativity (GR) and Teleparallel Gravity (TPG) are conceptually equivalent two theories, there are some fundamental conceptual differences between them. These differences make TPG to be a suitable framework for addressing some challenging problems in GR. The most important difference is that in TPG a tetrad field, instead of a metric in GR, is the fundamental geometric object.
In TPG the gravitation is described by torsion, not through geometrization as in
GR\footnote{In GR, gravity curves the space-time and shapes the geometry,
that is, the curvature geometries the gravitational interaction.}, as a
force acting on a test particle, that is, there are no geodesic equations \cite{HS79}.\\
In this paper, we restrict our attention to study the torsion vector, torsion axial-vector and gravitational energy
in the teleparallel version of Lewis-Papapetrou space-time.\\
From the geometric and physical view points the torsion axial-vector \cite{HS79}-\cite{OV04}
\begin{enumerate}
\item describes the deviation of the axial symmetry from spherical symmetry,\\
\item represents the spin precession of a Dirac particle,
\item is the gravitomagnetic part of the gravitational field and it is responsible for the Lense-Thirring effect, in the case of slow-rotation and weak field approximations,
\item is related to the space-time vorticity and its value is proportional to the magnitude of vorticity.\\
These effects can be explained by the non-invariance of Teleparallel Gravity under local transformations $SO(3)$.
\end{enumerate}
\par
In the framework  of Teleparallel Gravity,  the torsion axial vector
have studied by many authors   some of them have shown
that the spin precession of a Dirac particle is intimately related
to the torsion  axial-vector \cite{{Gad12},{PVZ},{Z05},{CLS03},{ZB02},{A81}}.
\par
The issue of distribution of energy-momentum is one of the oldest interesting and challenging problems, since the beginning of general relativity. In attempts to obtain an acceptable general definition of the gravitational field in this issue, different definitions were introduced \cite{EBM}.
These definitions are restricted to calculate the energy-momentum distribution in quasi-Cartesian coordinates, except M{\o}ller definition
\cite{E1}, in order to get a meaningful result.
\par
Although there is still no generally accepted definition for the gravitational energy-momentum distribution, the definitions given have yielded acceptable results. For a given space-time, some interesting results lead
to the conclusion that these prescriptions give the same
energy-momentum distribution \cite{Esame}. However, some examples
of space-times have been explored which do not support these
results \cite{Enot}.
\par
The theory of Teleparallel Gravity reproduces the empirical
content of the theory of General Relativity, but in a format that
more closely resembles the gauge theories of the standard model
than gravity does. Therefore, this theory allows a more
satisfactory treatment of the problem of energy-momentum
localization \cite{M62}. By working in the context of Teleparallel
Gravity, there are many attempts to find an acceptable definition
of the energy-momentum distribution for a given space-time. The
first attempt, after Einstein proposed this theory, was by
M{\o}ller \cite{M62}, who noticed that the tetrad description of
the gravitational field overcomes the problem of the gravitational
energy-momentum.
\par
This paper is organized  in the following way:  In the next
section we outline the general formalism of the fundamental
concepts of the theory of Teleparallel Gravity and
present the correct form of the inverse of tetrad field, given in \cite{SA06}, for the space-time under consideration
and obtain the correct expressions for the torsion vector and torsion
axial-vector. Moreover, we choose  another  set of tetrad field different from that given in \cite{SA06}. This tetrad satisfies the tetrad's conditions and leads to the teleparallel version of Lewis-Papapetrou space-time. We see if the physical and geometric objects of this space-time depend on the choice of tetrad or not.  In section 3, we calculate the energy and momentum
densities for the space-time under consideration. Finally, in section 4,  discussion and
conclusions are presented.

\setcounter{equation}{0}
\section{Teleparallel version of the Lewis-Papapetrou space-time}
We consider the stationary axisymmetric space-time represented by the following Lewis-Papapetrou line element \cite{SKM03}
\begin{equation}\label{metric}
ds^2=e^{2\psi}(dt-\omega d\theta)^2 - e^{2(\gamma - \psi)}(d\rho^2+dz^2)-\rho^2e^{-2\psi}d\theta^2,
\end{equation}
where $\gamma$, $\psi$ and $\omega$ are arbitrary functions of $\rho$ and $z$ only and $\omega$ represents the angular velocity. These functions satisfy the following constraint equations
\begin{equation}\label{psi}
\ddot{\psi}+\frac{1}{\rho}\dot{\psi}+\psi^{\prime\prime}=0,
\end{equation}
\begin{equation}\label{gamma}
\dot{\gamma}=\rho (\dot{\psi}^2-\psi^{\prime 2}), \qquad \gamma^\prime =2\rho\psi^\prime\dot{\psi}.
\end{equation}

As we stated in the introduction, in the context of TPG the tetrad field $h^{a}_{\,\,\,\mu}$ is used instead of the metric tensor $g_{\mu\nu}$. A given metric tensor admits many tetrad field. There are many ways to construct the tetrad $h^{a}_{\,\,\,\mu}$ which corresponds to $g_{\mu\nu}$. One of these ways is by solving the following relation
\begin{equation}\label{g}
g_{\mu\nu}=\eta_{ab}h^a_{\,\,\mu}h^b_{\,\,\nu},
\end{equation}
where $\eta_{ab}= diag (1, -1, -1, -1)$.\footnote{Here the Greek indices
$\mu, \nu, \lambda, ...$ have the range $0, ..., 3,$ and denote to tensor
indices, i.e. indices related to space-time; Latin indices $a, b, c, ...,$
have the range $0, ..., 3,$  will be used to denote local Lorentz (or tangent space) indices.}
A non-trivial tetrad $h^a_{\,\,\mu}$ and its inverse $h_a^{\,\,\nu}$ satisfy the following relations
\begin{equation}\label{h}
h^a_{\,\,\mu}h_a^{\,\,\nu} = \delta_{\mu}^{\nu}; \quad  h^a_{\,\,\mu}h_b^{\,\,\mu} = \delta_{b}^{a},
\end{equation}
and, in addition, the root of the metric determinant is
\begin{equation}\label{h-g}
h=\det (h^a_{\,\,\mu})=\sqrt{-g},
\end{equation}
where $g=\det( g_{\mu\nu})$.

The parallel transport of the tetrad $h^{a}_{\,\, \nu}$ between
two neighboring points is encoded in the covariant derivative
$$
\nabla_{\mu} h^{a}_{\,\, \nu} =\partial_{\mu} h^{a}_{\,\, \nu} -
\Gamma^{\alpha}_{\mu\nu}h^{a}_{\,\, \alpha},
$$
where
\begin{equation}\label{conn}
\Gamma^{\alpha}_{\,\,\mu\nu}= h_{a}^{\,\, \alpha}\partial_{\nu}
h^{a}_{\,\, \mu}=-h^{a}_{\,\, \mu}\partial_{\nu}h_{a}^{\,\, \alpha}
\end{equation}
is the Weitzenb\"{o}ck connection. This connection is presenting
torsion, but no curvature. The torsion of the Weitzenb\"{o}ck
connection is defined by
\begin{equation}\label{tor}
T^{\rho}_{\,\,\mu\nu}=\Gamma^{\rho}_{\,\,\nu\mu} -
\Gamma^{\rho}_{\,\,\mu\nu}=h_{a}^{\,\, \alpha}(\partial_{\mu}
h^{a}_{\,\, \nu}-\partial_{\nu} h^{a}_{\,\, \mu} )
\end{equation}
which is antisymmetric in the two indices $\mu$ and $\nu$.\\
The connection $\Gamma^{\rho}_{\,\, \mu\nu}$ and the Levi-Civita
connection $\tilde{\Gamma}^{\rho}_{\,\, \mu\nu}$\footnote{
$\tilde{\Gamma}^{\rho}_{\,\, \mu\nu}
=\frac{1}{2}g^{\rho\sigma}(g_{\mu\sigma,\nu}+g_{\nu\sigma,\mu}
-g_{\mu\nu,\sigma})$} are related by the following relation
$$
\Gamma^{\rho}_{\,\, \mu\nu}=\tilde{\Gamma}^{\rho}_{\,\, \mu\nu}+
K^{\rho}_{\,\, \mu\nu},
$$
where
$$
K^{\rho}_{\,\, \mu\nu}= \frac{1}{2}\big(T^{\,\,\rho}_{\mu\,\,
\nu}+T^{\,\,\rho}_{\nu\,\, \mu}- T^{\rho}_{\,\, \mu\nu}\big)
$$
is the {\it{contortion tensor}}.\\
The torsion tensor $T_{\lambda\mu\nu}$ can be decomposed into three
irreducible parts under the group of global Lorentz transformation
as follows \cite{HS79}
\begin{equation}
T_{\lambda\mu\nu}= \frac{1}{2}\big( t_{\lambda\mu\nu}
-t_{\lambda\nu\mu} \big)+
 \frac{1}{3}\big(g_{\lambda\mu}V_{\nu} - g_{\lambda\nu}V_{\mu} \big) + \varepsilon_{\lambda\mu\nu\rho}A^{\rho},
\end{equation}
where
$$
t_{\lambda\mu\nu}=\frac{1}{2}\big(T_{\lambda\mu\nu}+T_{\mu\lambda\nu}\big)
+\frac{1}{6}\big(g_{\nu\lambda}V_{\mu}+g_{\mu\nu}V_{\lambda}\big)
-\frac{1}{3}g_{\lambda\mu}V_{\nu}
$$
is the tensor part that defines the torsion tensor,
\begin{equation}\label{V}
V_{\mu}=T^{\nu}_{\,\,\nu\mu}
\end{equation}
is the vector part that defines  the torsion vector, and
\begin{equation} \label{ax}
A^{\mu}=h_{a}^{\,\,\mu}A^{a}=\frac{1}{6}\varepsilon^{\mu\nu\rho\sigma}T_{\nu\rho\sigma}
\end{equation}
is the axial-vector part that defines  the torsion axial-vector,
which represents
the deviation of the axial symmetry from spherical symmetry \cite{NH80}.\\
Here the completely antisymmetric tensors
$\varepsilon^{\mu\nu\rho\sigma}$ and
$\varepsilon_{\mu\nu\rho\sigma}$ with respect to the coordinates
basis are defined by \cite{M52}
$$
\varepsilon^{\mu\nu\rho\sigma}=\frac{1}{\sqrt{-g}}\delta^{\mu\nu\rho\sigma},
$$
$$
\varepsilon_{\mu\nu\rho\sigma}=\sqrt{-g}\delta_{\mu\nu\rho\sigma},
$$
where $\delta^{\mu\nu\rho\sigma}$ and $\delta_{\mu\nu\rho\sigma}$
are the completely antisymmetric tensor densities of weight $-1$ and
$+1$, respectively, with normalization $\delta^{0123}=+1$ and
$\delta_{0123}=-1$.\\

 In the context of TPG, it has been shown that the spin precession of a Dirac particle is intimately related to the torsion axial-vector via the vector differential equations \cite{HS79,NH80,YS80}.
\begin{equation}\label{Dirac}
\frac{d{\bf{S}}}{dt} = {\bf{\Omega}}\times {\bf{S}},
\end{equation}
where ${\bf{S}}$ is the spin vector of a Dirac particle and ${\bf{\Omega}}=-\frac{3}{2}{\bf{A}}$ is the Lense-Thirring precession angular velocity \cite{HN90}-\cite{CW95} with ${\bf{A}}$ to represent the space-like part of the torsion axial-vector. In GR, ${\bf{\Omega}}$ is produced by gravitomagnetic component of the gravitational field \cite{CW95}. The spin-rotation can be extended to gravitomagnetism via the Hamiltonian
\begin{equation}\label{Ham}
\delta H =  {\bf{\Omega}}\cdot {\bf{\sigma}},
\end{equation}
where ${\bf{\sigma}}$ is the particle spin.\\

The action of Teleparallel Gravity in the presence of matter is
given by

$$
{\A }= {\frac{1}{16\pi}} \int d{^4}x h S^{\rho\mu\nu}T_{\rho\mu\nu}+\int
d^4x h \pounds_{M},
$$
where $h=det(h^a_{\,\,\mu})$, $\pounds_{M}$ is the Lagrangian of a
source field and
\begin{equation}\label{S}
S^{\rho\mu\nu}= c_{1}T^{\rho\mu\nu} +
\frac{c_{2}}{2}\big(T^{\mu\rho\nu} - T^{\nu\rho\mu}\big) +
\frac{c_{3}}{2}\big(g^{\rho\nu}T^{\sigma\mu}_{\,\,\,\,\sigma} -
g^{\mu\rho}T^{\sigma\nu}_{\,\,\,\,\sigma}\big).
\end{equation}
is a tensor written in terms of the torsion of the
Weitzenb\"{o}ck connection. In the above form $c_{1}, c_{2}$ and
$c_{3}$ are the three
dimensionless coupling constants of Teleparallel Gravity.\\
For the so called teleparallel equivalent of General Relativity,
the specific choice of these constants is given by \cite{HS79}
\begin{equation}\label{2.13}
c_{1} =\frac{1}{4}, \qquad c_{2}=\frac{1}{2}, \qquad c_{3}=-1.
\end{equation}
The energy-momentum complexes of Einstein and
Landau-Lifshitz in Teleparallel Gravity, respectively, are given
by \cite{{M62}}
\begin{equation}\label{EBL}
\begin{array}{lcl}
hE^{\mu}_{\,\,\,\,\nu} & =
\frac{1}{4\pi}\partial_{\lambda}\Big(\mho_{\nu}^{\,\,\mu\lambda}
\Big),\\
hL^{\mu\nu} & =
\frac{1}{4\pi}\partial_{\lambda}\Big(hg^{\mu\beta}\mho_{\beta}^{\,\,\nu\lambda}\Big),
\end{array}
\end{equation}
where $\mho_{\nu}^{\,\,\mu\lambda}$ is the Freud's super-potential
and defined as follows
\begin{equation}\label{U}
\mho_{\nu}^{\,\,\mu\lambda}=hS_{\nu}^{\,\,\mu\lambda}.
\end{equation}
 The energy and momentum distributions in
the above complexes, respectively, are
\begin{equation}\label{e-mD}
\begin{array}{lcl}
P^{E}_\mu & = \int_{\Sigma}hE^0_{\,\,\mu}d^3x,\\
P^{LL}_\mu & = \int_{\Sigma}hL^0_{\,\,\mu}d^3x,
\end{array}
\end{equation}
where $P_{0}$ is the energy, $P_{i}\quad (i=1,2,3)$ are the
momentum components and the integration hypersurface $\Sigma$ is
described by $x^0 =t$ constant.\\

\par
 In the tetrad field for the space-time (\ref{metric}) given in \cite{SA06} (see equation (42) in \cite{SA06}), the authors considered the angle $\theta$ appeared in this tetrad represents the coordinate $x_2=\theta$. It is interested to point out that this angle is any constant angle satisfied the Pythagorean identity $\cos^2\theta+\sin^2\theta=1$ and it can be replaced by another symbol like $\Theta$,  which is different from the $x_2$-coordinate. Since the tetrad and its inverse given in \cite{SA06} (see equations (42) and (43) in \cite{SA06}) do not satisfy the tetrad's conditions (\ref{g}) and (\ref{h}). Therefore, we will re-writing the tetrad and its inverse   on the right form, to satisfied the relations (\ref{g}) and (\ref{h}),  as follows
 \begin{equation}\label{T}
{h^a_{\,\,\nu}} =
\left[ \begin{array}{cccc}
e^{-\psi} & 0 & -\omega e^{\psi} & 0 \\
0 & e^{\gamma-\psi}\cos\Theta & -\rho e^{-\psi}\sin\Theta& 0 \\
0 & e^{\gamma-\psi}\sin\Theta & \rho e^{-\psi}\cos\Theta & 0 \\
0 & 0 & 0 & e^{\gamma-\psi}
\end{array} \right],
\end{equation}

\begin{equation}\label{IT}
{h_a^{\,\,\nu}} =
\left[ \begin{array}{cccc}
e^{-\psi} & -\omega\rho^{-1}e^{\psi}\sin\Theta & \omega\rho^{-1}e^{\psi}\cos\Theta & 0 \\
0 & e^{-\gamma+\psi}\cos\Theta & e^{-\gamma+\psi}\sin\Theta & 0 \\
0 & -\rho^{-1}e^{\psi}\sin\Theta& \rho^{-1}e^{\psi}\cos\Theta & 0 \\
0 & 0 & 0 & e^{-\gamma+\psi}
\end{array} \right].
\end{equation}
According to previous modifications, the components of the Weitzenb\"{o}ck connection, $\Gamma^\sigma_{\,\,\mu\nu}$, vanish when $\nu=2$. Therefore the components $\Gamma^0_{\,\,12}$, $\Gamma^1_{\,\,22}$ and $\Gamma^2_{\,\,12}$ obtained in \cite{SA06} (see equation (44) in \cite{SA06}) should identically be equal zero. This leads to  correct the components $T^0_{\,\,12}$ and $T^2_{\,\,12}$ of the torsion tensor $T^\sigma_{\,\,\mu\nu}$ (see equation (45) in \cite{SA06}),  to become
\begin{equation}\label{t}
\begin{array}{ccc}
T^0_{\,\,12} &=-T^0_{\,\,21}&=\omega\rho^{-1}-(\dot{\omega}+2\omega\dot{\psi}),\\
T^2_{\,\,12} &=-T^2_{\,\,21}&=\rho^{-1}(1-\rho\dot{\psi}).
\end{array}
\end{equation}
The another non-vanishing components of the torsion tensor $T^\sigma_{\,\,\mu\nu}$ (see equation (45) in \cite{SA06}), are
\begin{equation}\label{SATT}
\begin{array}{clcl}
T^0_{\,\,01}= & -\dot{\psi},\\
T^0_{\,\,03}= & -\psi^\prime,\\
T^0_{\,\,23}= & \omega^\prime+2\omega\psi^\prime,\\
T^1_{\,\,13}= & \psi^\prime-\gamma^\prime,\\
T^2_{\,\,23}= & \psi^\prime,\\
T^3_{\,\,31}=& \dot{\psi}-\dot{\gamma}.
\end{array}
\end{equation}

Consequently to the above corrections, the non-vanishing components of the  torsion vector become
\begin{equation}\label{tv}
\begin{array}{ccc}
v_1&=\dot{\psi}-\dot{\gamma}-\frac{1}{\rho},\\
v_3&=\psi^\prime-\gamma^\prime,
\end{array}
\end{equation}
and the space-like axial-vector given in \cite{SA06} will now completely improved to the following form
\begin{equation}\label{av}
{\bf{A}}=-\frac{1}{3\rho}e^{2(2\psi-\gamma)}[(\omega^\prime+2\omega\psi^\prime)\hat{e}_\rho-\dot{\omega}\hat{e}_z],
\end{equation}
where $\hat{e}_\rho$ and $\hat{e}_z$ are the unit vectors along radial and $z$-directions, respectively.
Here dot and prime denote the derivative with respect to $\rho$ and $z$, respectively.\\
The expression (\ref{av}) shows that the space-like part of the torsion axial-vector lies in the $(-\rho)z$-plane.\\
We note that $h=\sqrt{-g}=\rho e^{2(\gamma-\psi)}$, not as appeared in \cite{SA06}.
\\
After giving the correct expression of the torsion axial-vector, we will give the expressions of the spin procession of the Dirac particle and the corresponding Hamiltonian, respectively, as follows
\begin{equation}\label{Dirac0}
\frac{d{\bf{S}}}{dt} =\frac{e^{2(2\psi-\gamma)}}{2\rho} [(\omega^\prime+2\omega\psi^\prime)\hat{e}_\rho-\dot{\omega}\hat{e}_z] \times {\bf{S}},
\end{equation}
\begin{equation}\label{Ham0}
\delta {\bf{H}}=   \frac{e^{2(2\psi-\gamma)}}{2\rho} [(\omega^\prime+2\omega\psi^\prime)\hat{e}_\rho-\dot{\omega}\hat{e}_z]\cdot {\bf{\sigma}},
\end{equation}

Now, we will correct the depended results on special cases of the values of the functions $\omega$ (see equations (52)-(54) in \cite{SA06})
\begin{enumerate}
\item if $\omega$ depends only on $z$, then the axial-vector will be symmetric about radial axis and takes the form
\begin{equation}
{\bf{A}}=-\frac{1}{3\rho}e^{2(2\psi-\gamma)}[(\omega^\prime+2\omega\psi^\prime)\hat{e}_\rho].
\end{equation}
The expressions of $\frac{d{\bf{S}}}{dt}$ and $\delta {\bf{H}}$ are
$$
\frac{d{\bf{S}}}{dt} =\frac{e^{2(2\psi-\gamma)}}{2\rho} [(\omega^\prime+2\omega\psi^\prime)\hat{e}_\rho] \times {\bf{S}},
$$
$$
 \delta {\bf{H}}=   \frac{e^{2(2\psi-\gamma)}}{2\rho} [(\omega^\prime+2\omega\psi^\prime)\hat{e}_\rho]\cdot {\bf{\sigma}}.
 $$
\item if $\omega$ is only a function of $\rho$, then expressions of ${\bf{A}}$, lies in $\rho z$-plane, $\frac{d{\bf{S}}}{dt}$ and $\delta {\bf{H}}$ are, respectively
$$
{\bf{A}}=-\frac{1}{3\rho}e^{2(2\psi-\gamma)}[(2\omega\psi^\prime)\hat{e}_\rho-\dot{\omega}\hat{e}_z].
$$
$$
\frac{d{\bf{S}}}{dt} =\frac{e^{2(2\psi-\gamma)}}{2\rho} [(2\omega\psi^\prime)\hat{e}_\rho-\dot{\omega}\hat{e}_z] \times {\bf{S}},
$$
$$
 \delta {\bf{H}}=   \frac{e^{2(2\psi-\gamma)}}{2\rho} [(2\omega\psi^\prime)\hat{e}_\rho-\dot{\omega}\hat{e}_z]\cdot {\bf{\sigma}}.
 $$
\item if $\omega$ is constant, then
$$
{\bf{A}}=-\frac{1}{3\rho}e^{2(2\psi-\gamma)}[(2\omega\psi^\prime)\hat{e}_\rho].
$$
$$
\frac{d{\bf{S}}}{dt} =\frac{e^{2(2\psi-\gamma)}}{2\rho} [(2\omega\psi^\prime)\hat{e}_\rho] \times {\bf{S}},
$$
$$
 \delta {\bf{H}}=   \frac{e^{2(2\psi-\gamma)}}{2\rho} [(2\omega\psi^\prime)\hat{e}_\rho]\cdot {\bf{\sigma}}.
 $$
\end{enumerate}
It is interested to point out that these corrections will correct the geometrical and
physical meaning of the results obtained by Sharif and Amir \cite{SA06} and give the true
discussion of the geometry and physics of the space-time under consideration.

\par
Now, we will construct   another tetrad field related to the metric
(\ref{metric})  to be different  from the one given in (\ref{T}), and
to be of the form
\begin{equation}\label{ST}
{\tilde{h}^a_{\,\,\mu}} =
\left[ \begin{array}{cccc}
e^{\psi} & 0 & -\omega e^{\psi} & 0 \\
0 & e^{\gamma-\psi} & 0 & 0 \\
0 & 0 & \rho e^{-\psi} & 0 \\
0 & 0 & 0 & e^{\gamma-\psi}
\end{array} \right],
\end{equation}

 its inverse is
\begin{equation}\label{IST}
{\tilde{h}_a^{\,\,\mu}} =
\left[ \begin{array}{cccc}
e^{-\psi} & 0 & \frac{\omega e^\psi}{\rho} & 0 \\
0 & e^{\psi-\gamma} & 0 & 0 \\
0 & 0 & \frac{e^\psi}{\rho} & 0 \\
0 & 0 & 0 & e^{\psi-\gamma}
\end{array} \right].
\end{equation}
We can inspect that equations (\ref{ST}) and (\ref{IST}) satisfy the tetrad's conditions (\ref{g}) and (\ref{h})
and $\tilde{h}=\det \tilde{h}^a_{\,\,\mu}=\sqrt{-g}=\rho e^{2(\gamma-\psi)}$.\\
Using  this tetrad field and its inverse in equations (\ref{conn}),
 we get the non-vanishing components of the Weitzenb\"{o}ck connections

\begin{equation}\label{W-C}
\begin{array}{clcl}
\tilde{\Gamma}^0_{\,\,01}= & \dot{\psi},\\
\tilde{\Gamma}^0_{\,\,03}= & \psi^\prime,\\
\tilde{\Gamma}^0_{\,\,21}= & \frac{\omega}{\rho}-(\dot{\omega}+2\omega\dot{\psi}),\\
\tilde{\Gamma}^0_{\,\,23}= & -(\omega^\prime+2\omega\psi^\prime),\\
\tilde{\Gamma}^1_{\,\,11}= & \dot{\gamma}-\dot{\psi}=\Gamma^3_{\,\,31},\\
\tilde{\Gamma}^1_{\,\,13}= & \gamma^\prime-\psi^\prime=\Gamma^3_{\,\,33},\\

\tilde{\Gamma}^2_{\,\,21}= & \frac{1}{\rho}-\dot{\psi},\\
\tilde{\Gamma}^2_{\,\,23}= & -\psi^\prime.\\

\end{array}
\end{equation}

 Inserting the above components into (\ref{tor}), we obtain the corresponding non-vanishing components of the torsion tensor

\begin{equation}\label{FTT}
\begin{array}{clcl}
\tilde{T}^0_{\,\,01}= & -\dot{\psi},\\
\tilde{T}^0_{\,\,03}= & -\psi^\prime,\\
\tilde{T}^0_{\,\,21}= & -\frac{\omega}{\rho}+(\dot{\omega}+2\omega\dot{\psi}),\\
\tilde{T}^0_{\,\,23}= & \omega^\prime+2\omega\psi^\prime,\\

\tilde{T}^1_{\,\,13}= & \psi^\prime-\gamma^\prime,\\

\tilde{T}^2_{\,\,21}= & \dot{\psi}-\frac{1}{\rho},\\
\tilde{T}^2_{\,\,23}= & \psi^\prime,\\
\tilde{T}^3_{\,\,31}=& \dot{\psi}-\dot{\gamma}.
\end{array}
\end{equation}
 Substituting the  above components in equations (\ref{V}) and (\ref{ax}), we get, respectively,
  the non-vanishing components of the torsion vector $V_\mu$ and  the torsion axial-vector $A^\mu$ as follows

\begin{equation}\label{Sv}
\begin{array}{lcl}
\tilde{v}_1= \dot{\psi}-\dot{\gamma}-\frac{1}{\rho},\\
\tilde{v}_3 =\psi^\prime-\gamma^\prime,
\end{array}
\end{equation}
\begin{equation}\label{Sax-v}
\begin{array}{lcl}
\tilde{A}^1=-\frac{e^{2\psi}}{3h}[\omega^\prime+2\omega\psi^\prime]=-\frac{e^{2(\gamma+2\psi)}}{3\rho}[\omega^\prime+2\omega\psi^\prime]\\
\tilde{A}^3 = \frac{e^{2\psi}}{3h}\dot{\omega}=\frac{e^{2(2\psi-\gamma)}}{3\rho}\dot{\omega}.
\end{array}
\end{equation}
The space-like torsion axial-vector has the following vector form
\begin{equation}\label{A2}
{\bf{\tilde{A}}}=-\frac{e^{2(2\psi-\gamma)}}{3\rho} [(\omega^\prime+2\omega\psi^\prime)\hat{e}_\rho-\dot{\omega}\hat{e}_z].
\end{equation}
This expression shows that the torsion axial-vector will be
lies in the $(-\rho)z$-plane.\\
Equations (\ref{tv}) and (\ref{Sv})  and equations (\ref{av}) and  (\ref{A2}) show, respectively, that both torsion vector and axial-vector do not depend on the choice of tetrad field and have the same values in both two sets of tetrad field.\\

Using equation (\ref{A2}) in equations (\ref{Dirac}) and (\ref{Ham}), we obtain  the spin procession of the Dirac particle and the corresponding Hamiltonian, respectively
\begin{equation}\label{Dirac1}
\frac{d{\bf{\tilde{S}}}}{dt} =\frac{e^{2(2\psi-\gamma)}}{2\rho} [(\omega^\prime+2\omega\psi^\prime)\hat{e}_\rho-\dot{\omega}\hat{e}_z] \times {\bf{S}},
\end{equation}
\begin{equation}\label{Ham1}
\delta {\bf{\tilde{H}}}=   \frac{e^{2(2\psi-\gamma)}}{2\rho} [(\omega^\prime+2\omega\psi^\prime)\hat{e}_\rho-\dot{\omega}\hat{e}_z]\cdot {\bf{\sigma}}.
\end{equation}
The above two equations are the same equations (\ref{Dirac0}) and (\ref{Ham0}) obtained using the tetrad field given in \cite{SA06}.

\section{Energy and momentum densities}

This section deals with calculating  the energy and momentum densities of
the space-time (\ref{metric}) in the context of the theory of
teleparallel gravity. \\
 Inserting the components (\ref{FTT}) to (\ref{S}), using (\ref{2.13}),
 we obtain the following non-vanishing required components of $S_\beta^{\,\,\mu \nu}$
\begin{equation}
\begin{array}{cccc}
 S_0^{\,\,01} &= & \frac{1}{4}e^{2(\psi-\gamma)}\Big[ \frac{\omega\dot{\omega}}{\rho^2}e^{4\psi}-\frac{2}{\rho}-2\dot{\gamma}+4\dot{\psi}\Big],\\
 S_0^{\,\,03}& = & \frac{1}{4}e^{2(\psi-\gamma)}\Big[ \frac{\omega\omega^\prime}{\rho^2}e^{4\psi}-2\gamma^\prime+4\psi^\prime\Big],\\
  S_0^{\,\,12}&=  &-e^{2(\psi-\gamma)}\Big[\frac{\dot{\omega}}{4\rho^2}e^{4\psi}\Big],\\
  S_0^{\,\,23} &=  &e^{2(\psi-\gamma)}\Big[\frac{\omega^\prime}{4\rho^2}e^{4\psi}\Big],\\
 S_2^{\,\,01} &=  &-\frac{1}{4}e^{2(\psi-\gamma)}\Big[ \frac{\omega^2\dot{\omega}}{\rho^2}e^{4\psi}-\frac{2\omega}{\rho}+\dot{\omega}+4\omega\dot{\psi}\Big],\\
 S_2^{\,\,03}& = & -\frac{1}{4}e^{2(\psi-\gamma)}\Big[ \frac{\omega^2\omega^\prime}{\rho^2}e^{4\psi}+\omega^\prime+4\omega\psi^\prime\Big],\\
  S_2^{\,\,12}& = &- e^{2(\psi-\gamma)}\Big[\frac{\omega\dot{\omega}}{4\rho^2}e^{4\psi}+\frac{\dot{\gamma}}{2}\Big],\\
  S_2^{\,\,23}& = & -e^{2(\psi-\gamma)}\Big[\frac{\omega\omega^\prime}{4\rho^2}e^{4\psi}+\frac{\gamma^\prime}{2}\Big].
\end{array}
\end{equation}

Inserting  these components  to equations (\ref{U}) and
(\ref{EBL}), using the relations (\ref{psi}) and (\ref{gamma}),  we get the energy and momentum densities in the
prescriptions of Einstein, and Landau-Lifshitz,
respectively, as follows:
\begin{equation}\label{Einstein}
\begin{array}{ccc}
hE^{0}_{0}& =\frac{e^{4\psi}}{16\pi\rho^2}\Big[4\rho\omega(\dot{\omega}\dot{\psi}+\omega^\prime\psi^\prime)+
\rho(\dot{\omega}^2+\omega^{\prime 2})-\omega\dot{\omega}+\omega\ddot{\omega}+\omega^{\prime\prime}+
4\rho^3\psi^{\prime 2}e^{-4\psi}\Big],\\
hE^{0}_{2}& =-\frac{1}{16\pi}\Big[\big(\rho+\frac{\omega^2}{\rho}e^{4\psi}\big)\big(-\frac{\dot{\omega}}{\rho}+
4(\dot{\omega}\dot{\psi}+\omega^\prime\psi^\prime)+\ddot{\omega}+\omega^{\prime\prime}\big)+\frac{2\omega e^{4\psi}}{\rho}(\dot{\omega}^2+\omega^{\prime 2})\Big],\\
& hE^{0}_{1} =hE^{0}_{3} =0.
\end{array}
\end{equation}
\begin{equation}\label{Landau}
\begin{array}{ccc}
hL^{00}& =\frac{e^{2\gamma-4\psi}}{8\pi}\Big[ -1  +6\rho\dot{\psi}-3\rho\dot{\gamma}
-2\rho^2(4\dot{\psi}^2+3\psi^{\prime 2}+\dot{\gamma}^2+\gamma^{\prime 2} + 4\dot{\psi}\dot{\gamma}+4\psi^\prime\gamma^\prime)+\\
&e^{4\psi}\Big(\omega(4\dot{\omega}\dot{\gamma}+4\omega^\prime\gamma^\prime+\ddot{\omega}+\omega^{\prime\prime})+ \dot{\omega}^2 +\omega^{\prime 2} +\omega^2(\ddot{\gamma}+\gamma^{\prime\prime}+2\dot{\gamma}^2+2\gamma^{\prime 2})\Big)\Big],\\
hL^{02}& =\frac{\omega e^{2\gamma+4\psi}}{4\pi\rho^2}\Big[\dot{\omega}^2+\frac{\omega\ddot{\omega}}{2}+2\omega\dot{\omega}\dot{\psi}+
\omega\dot{\omega}\dot{\gamma}-\frac{\omega\dot{\omega}}{\rho}+\\
&\rho^2e^{-4\psi}\big(\dot{\gamma}^2+\gamma^{\prime 2}+\frac{1}{2}(\ddot{\gamma}+\gamma^{\prime\prime})+\frac{1}{\omega}(\omega^\prime\gamma^\prime+\frac{\omega^{\prime\prime}}{4}-\frac{\ddot{\omega}}{4})\big)\Big],\\
& hL^{01} =hL^{03} =0.
\end{array}
\end{equation}

These results are obtained by using the tetrad given in (\ref{T}) (see equation (42) in \cite{SA06}) and its right inverse (\ref{IT}).\\

Using the other constructed tetrad (\ref{ST}) and its inverse (\ref{IST}), we get the same results for the torsion vector (\ref{tv}), space-like axial-vector (\ref{av}) and  energy and momentum densities in the sense of Einstein and Landau-Lifshitz (\ref{Einstein})-(\ref{Landau}).

The question that arises now is: Why did the two different tetrad fields give the same results to the aforementioned quantities?
We will postpone the answer to the conclusion.\\
We can conclude that if we choose two sets of relevant tetrad
fields, both of them achieve the tetrad's conditions, we find that
the
 energy and momentum densities do not depend on the choice of tetrad
 fields. This result sustain the result
obtained in \cite{M02}  that in the framework of the teleparallel
geometry the correct description of the gravitational
energy-momentum singles out a unique set of tetrad fields.

\section{Discussion and conclusion}
In this work we  focused on the  geometric and physical properties of
Lewis-Papapetrou space-time within the framework of the teleparallel gravity.\\
The results we have obtained here are different from that obtained in \cite{SA06} and \cite{SA07}.
Therefore, the analysis for the  torsion axial-vector,  energy-momentum density, spin precession of a Dirac particle and the extra
Hamiltonian of the space-time under consideration is completely new.\\
The second objective of this work was to create another set of tetrad field different from the one given in (\ref{T}) and to examine the geometric and physical quantities depend on  the choose of tetrad field or not. We found that the axial-vector, torsion axial-vector, the spin procession of the Dirac particle and the corresponding Hamiltonian do not depend on the choice of tetrad field.\\
The reason for this independence  came from the Lorentz transformation which links the two tetrad fields $h^a_{\,\,\mu}$ and $\tilde{h}^b_{\,\,\mu}$ by
$$
h^a_{\,\,\mu} =A^a_{\,\,b}\tilde{h}^b_{\,\,\mu},
$$
where the Lorentz factor, ${A^a_{\,\,b}}$, is given by
$$
{A^a_{\,\,b}} =
\left[ \begin{array}{cccc}
1 & 0 & 0 & 0 \\
0 & \cos\Theta & -\sin\Theta & 0 \\
0 & \sin\Theta & \cos\Theta & 0 \\
0 & 0 & 0 & 1
\end{array} \right].
$$
As it is well known, in the context of Riemannian geometry the space-time is invariant under the Lorentz transformation, therefore the Lorentz factor, ${A^a_{\,\,b}}$, is irrelevant. On the contrary,  in the context of Weitzenb\"{o}ck  geometry the Lorentz factor is the basic element of the definition of the tetrad field and, accordingly, the connection. This shows why the choice of tetrad field is a delicate issue in TPG.
For the space-time under study, we found that ${A^a_{\,\,b}}$ does not depend on the components of tetrad field, that is, it is irrelevant. Then the  teleparallel version of Lewis-Papapetrou space-time is invariant under Lorentz transformation. This  sustains that the theories of TPG and GR are  equivalent.\\
Gad \cite{Gad12} studied the teleparallel  version of Van Stockum space-time. He found that the obtained expressions for the torsion vector  and the torsion axial-vector are quite different for the two sets of tetrad fields. The reason  for this was that the Lorentz factor, given below\footnote{We corrected element $A^3_{\,\,3}$ after adding the forgotten amount $\ell^2$ in the numerator and abbreviation} (see \cite{Gad12}),  depends on the tetrad components

$$
{A^a_{\,\,b}} =
\left[ \begin{array}{cccc}
\sqrt{\frac{f\ell+m^2}{f\ell}} & 0 & 0 & -\frac{m}{\sqrt{f\ell}} \\
0 & 1 & 0 & 0 \\
0 & 0 & 1 & 0 \\
-\frac{m}{\sqrt{f\ell}}& 0 & 0 & \sqrt{\frac{f\ell+m^2}{f\ell}}
\end{array} \right].
$$

Now the question at the end of the previous section can be answered as follows:
As mentioned in reference 4, two tetrad fields associated with a local Lorentz transformation do not produce equivalent field equations associated with Lorentz transformation. We will add to this, that this can happen if the Lorentz factor depends on the tetrad components, $h_a^{\,\,\nu}$, (as in the case of Van Stockum space-time, see \cite{Gad12}). In the case of Lorentz's factor does not dependent on the tetrad components, the space-time is invariant under Lorentz transformation, as in the Riemannian geometry (as in the space-time under study).

\section*{Acknowledgments}
This project was funded by the Deanship of Scientific Research (DSR), King Abdulaziz University, Jeddah, under grant No. (130-1012-D1435).
The authors, therefore, acknowledge with thanks DSR technical and financial support.


\end{document}